\documentclass[manuscript]{aastex}

\usepackage{txfonts}
\usepackage{longtable}
\usepackage{rotating}
\usepackage{natbib}
\usepackage{graphicx}
\usepackage{graphics}
\usepackage{psfrag}
\usepackage{amssymb}
\bibpunct{(}{)}{;}{a}{}{,}
\def\Teff{$T_{\mathrm{eff}}$}
\def\logg{\ensuremath{\log g}}

\def\vsini{\ensuremath{{\upsilon}\sin i}}
\def\kms{$\mathrm{km\,s}^{-1}$}

\def\vr{${\upsilon}_{\mathrm{r}}$}

\def\espa{ESPaDOnS}

\def\llm{{\sc LLmodels}}

\def\synth{{\sc SYNTH3}}

\def\logR{\ensuremath{\log R^{'}_{HK}}}

\shorttitle{Absorbing gas around the WASP-12 planetary system}
\shortauthors{Fossati et al.}

\begin{document}


\title{Absorbing gas around the WASP-12 planetary system$^1$}

\altaffiltext{1}{Based on observations obtained at the Canada-France-Hawaii
Telescope (CFHT), which is operated by the National Research Council of
Canada, the Institut National des Sciences de l'Univers of the Centre
National de la Rechereche Scientifique of France, and the University of
Hawaii. Based on observations made with the Nordic Optical Telescope, operated on the island of La Palma jointly by Denmark, Finland, Iceland, Norway, and Sweden, in the Spanish Observatorio del Roque de los Muchachos of the Instituto de Astrofisica de Canarias.}








\author{L. Fossati\altaffilmark{2}}
\affil{Argelander-Institut f\"ur Astronomie der Universit\"at Bonn, Auf dem H\"ugel 71, 53121, Bonn, Germany}
\email{lfossati@astro.uni-bonn.de}
\and
\author{T.R. Ayres}
\affil{Center for Astrophysics and Space Astronomy, University of Colorado, 593 UCB, Boulder, CO 80309-0593, USA}
\email{Thomas.Ayres@colorado.edu}
\and
\author{C.A. Haswell}
\affil{Department of Physics and Astronomy, Open University,
	Walton Hall, Milton Keynes MK7 6AA, UK}
\email{C.A.Haswell@open.ac.uk}
\and
\author{D. Bohlender}
\affil{Herzberg Institute of Astrophysics, National Research Council of
	Canada, 5071 West Saanich Road, Victoria, BC V9E 2E7, Canada}
\email{david.bohlender@nrc-cnrc.gc.ca}
\and
\author{O. Kochukhov}
\affil{Department of Physics and Astronomy, Uppsala University, SE-751 20,
	Uppsala, Sweden}
\email{oleg.kochukhov@physics.uu.se}
\and
\author{L. Fl\"oer}
\affil{Argelander-Institut f\"ur Astronomie der Universit\"at Bonn, Auf dem H\"ugel 71, 53121, Bonn, Germany}
\email{lfloeer@astro.uni-bonn.de}

\altaffiltext{2}{Department of Physical Sciences, Open University, Walton Hall, Milton Keynes MK7 6AA, UK}


%
\begin{abstract}
Near-UV observations of the planet host star WASP-12 uncovered the apparent absence of the normally conspicuous core emission of the \ion{Mg}{2}\,h\&k resonance lines. This anomaly could be due either to (1) a lack of stellar activity, which would be unprecedented for a solar-like star of the imputed age of WASP-12; or (2) extrinsic absorption, from the intervening interstellar medium (ISM) or from material within the WASP-12 system itself, presumably ablated from the extreme hot Jupiter WASP-12\,b. HIRES archival spectra of the \ion{Ca}{2}\,H\&K lines of WASP-12 show broad depressions in the line cores, deeper than those of other inactive and similarly distant stars and similar to WASP-12's \ion{Mg}{2}\,h\&k line profiles. We took high resolution \espa\ and FIES spectra of three early-type stars within 20$\arcmin$ of WASP-12 and at similar distances, which show the ISM column is insufficient to produce the broad \ion{Ca}{2} depression observed in WASP-12. The EBHIS \ion{H}{1} column density map supports and strengthens this conclusion. Extrinsic absorption by material local to the WASP-12 system is therefore the most likely cause of the line core anomalies. Gas escaping from the heavily irradiated planet could form a stable and thick circumstellar disk/cloud. The anomalously low stellar activity index (\logR) of WASP-12 is evidently a direct consequence of the extra core absorption, so similar HK index deficiencies might signal the presence of translucent circumstellar gas around other stars hosting evaporating planets. 
\end{abstract}
%

\keywords{stars: individual (WASP-12) --- ISM: clouds --- Planets and satellites: individual (WASP-12) --- Planet-star interactions}

%
\section{Introduction}
WASP-12, a late F-type main sequence star younger than 2.65\,Gyr \citep{fossati2010b}, hosts one of the hottest, most bloated known exoplanets, orbiting extremely close to the host star \citep{hebb2009}. WASP-12\,b is one of the most irradiated transiting planets known, and provides key tests of the ``blow-off'' hypothesis whereby a significant fraction of the planet's atmosphere is being eroded and escapes.

\citet{fossati2010a} and \citet{haswell2012} analysed Hubble Space Telescope (HST) near-UV (NUV) spectra obtained with the Cosmic Origin Spectrograph (COS), confirming the on-going blow-off, similar to that of HD209458 \citep{vidal03,linsky2010}. NUV dips during the planetary crossings are up to three times deeper than the optical transits of WASP-12\,b, suggesting there is diffuse gas extending well beyond the planetary Roche lobe. \citet{li2010} suggested that the gas lost by the planet could form an accretion disk around the star. The HST data revealed a variable and early ingress, contrary to model expectations, but with egresses compatible with the optical ephemeris. The early-ingress phenomenon has been modeled by various authors \citep[e.g.,][]{lai2010,vidotto2010,bisikalo2013}, but these models do not yet explain the presence of an ingress as early as orbital phase $-$0.17 \citep{haswell2012}.

The HST data further disclosed a remarkable anomaly in the stellar NUV spectrum: the apparent absence of the normally bright emission cores in the \ion{Mg}{2}\,h\&k resonance lines \citep{haswell2012}. Significantly, the anomaly is always present, regardless of the planet's orbital phase. The absence of \ion{Mg}{2} core emission was completely unexpected given the spectral type and imputed age of the star \citep[e.g.,][]{guinan2009}. Figure\,16 of \citet{haswell2012} compares the near-UV spectra of WASP-12,  HD102634,  HD107213 and Procyon; Procyon is a mid-F subgiant commonly used as an inactive comparison star, while HD102634 and  HD107213 are similar in temperature and age \citep{valenti2005} to WASP-12. The comparison highlights the complete lack of emission in the WASP-12 line cores.

\citet{fossati2010b} and \citet{haswell2012} discussed scenarios that could lead to the anomalous line cores: (1) intrinsically low stellar activity; or (2) extrinsic absorption, either from the interstellar medium (ISM) or from material within the system, for example in a circumstellar disk.  \citet{haswell2012} showed that the ISM column density required to reproduce the apparent \ion{Mg}{2} absorption was 10 times greater than expected along the WASP-12 line of sight, but that planetary mass loss potentially could produce the large column required to extinct the \ion{Mg}{2} cores.

Opportunities to use the \ion{Mg}{2}\,h\&k lines to compare WASP-12 with other stars are few: inactive distant stars  have  rarely been observed at high resolution in the NUV, because of the bias of interest in NUV-bright stars, which either are active or nearby. The optical \ion{Ca}{2}\,H\&K resonance lines also display core emission of chromospheric origin closely related to that of the \ion{Mg}{2} lines, although not as strong \citep{hall2008}. The \ion{Ca}{2} lines have been used to assess the activity of extensive samples of late-type stars, including planet-hosts, some of which indeed are inactive and distant \citep{knutson2010}.

In Sect.~\ref{ca} we compare WASP-12's \ion{Ca}{2}\,H\&K line profiles with those of other inactive and distant planet-hosting stars; in Sect.~\ref{ism} we evaluate the ISM absorption along the WASP-12 line of sight; and in Sect.~\ref{discussion} we discuss possible origins of the anomalously depressed resonance line cores.
\section{The \ion{Ca}{2}\,H\&K line profiles: activity and anomaly}\label{ca}
It is well known that chromospheric activity correlates with the stellar rotational velocity. For WASP-12 the rotation rate is unknown, but the projected rotational velocity (\vsini) has an upper limit of only 1\,\kms\  (E. Simpson, private communication). Because a low \vsini\ could potentially imply an intrinsically low level of stellar activity, we must consider whether depressed chromospheric emission might be responsible for the \ion{Mg}{2} core anomaly. Ideally, we would compare with stars of similar type which do not host hot-Jupiters. However, planet-hosts are far better studied at high spectral resolution in the visible, so we exploited planet-host spectra obtained for planet detection through Doppler-reflex or stellar characterisation.

\begin{figure}
\includegraphics[width=11.5cm]{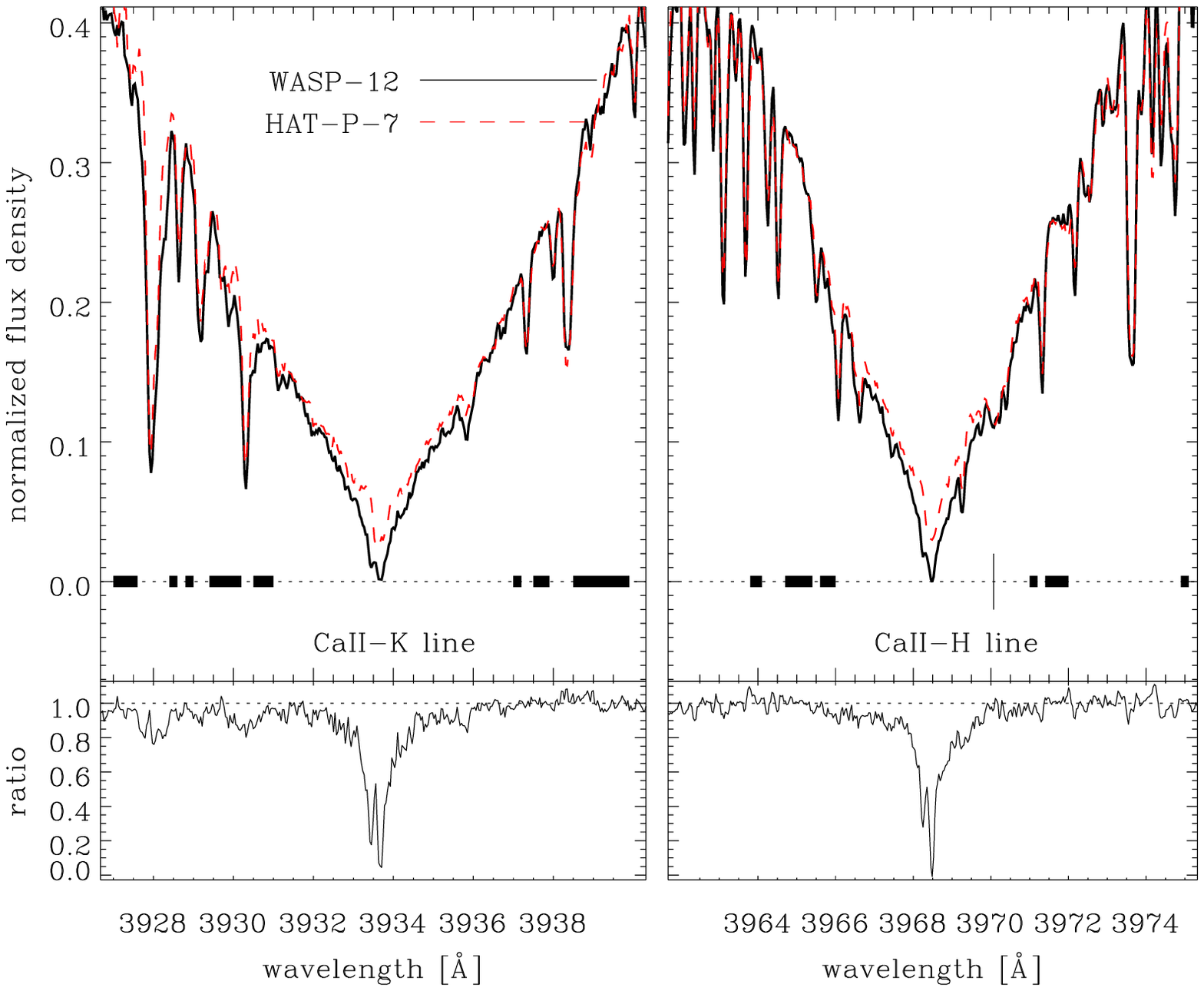}
\includegraphics[width=11.5cm]{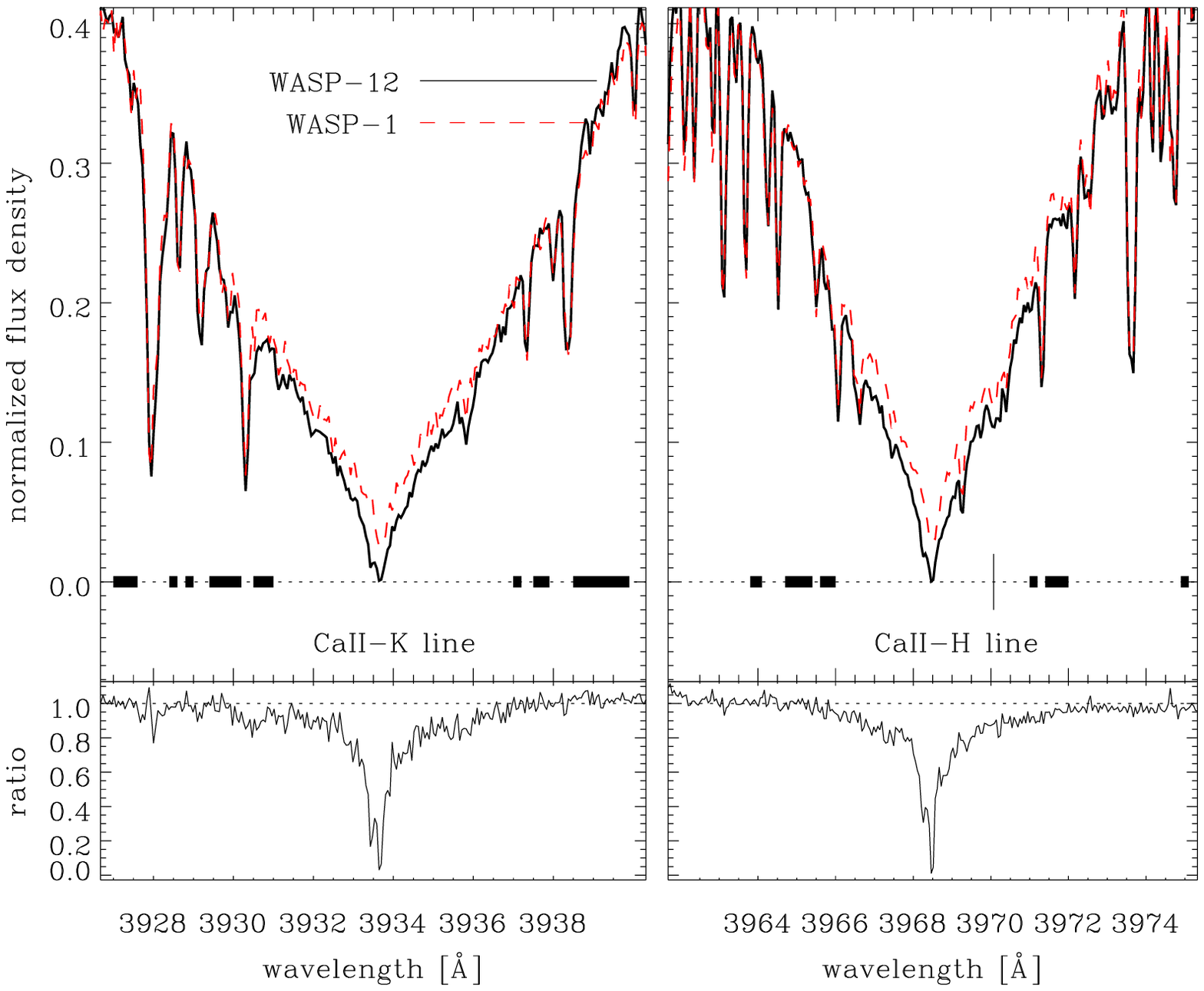}
\caption{\label{fig:corot1wasp1} Comparison between the \ion{Ca}{2}\,H\&K line profiles (right and left panels, respectively) of WASP-12 and HAT-P-7 (upper panels) or WASP-12 and WASP-1 (lower panels). Horizontal bars indicate spectral windows used to normalise the flux profiles. Short vertical lines in the right panels mark the location of hydrogen H$\epsilon$. The per pixel uncertainties are of the order of 3\%. The lower panel within each plot shows the ratio of the spectra with WASP-12's as the numerator. In each case WASP-12 has extra absorption throughout the inner $\sim$4\,\AA\ of the line profile.}
\end{figure}

Figure~\ref{fig:corot1wasp1} compares the WASP-12 \ion{Ca}{2}\,H\&K lines with those of HAT-P-7 and WASP-1. The spectra were acquired with the High Resolution Echelle Spectrometer (HIRES) on the KECK\,I telescope. We obtained the reduced data\footnote{\tt http://www.astro.caltech.edu/~tb/makee/} from the HIRES archive\footnote{\tt http://www2.keck.hawaii.edu/koa/public/koa.php}, and focused on spectra with the highest signal-to-noise ratio (S/N) and taken with the same instrument settings as the out-of-transit phases of WASP-12.

The fundamental parameters of the three stars are listed in Table~\ref{tab:comp}. Within the uncertainties, the comparison stars are comparable in effective temperature (\Teff) to WASP-12. Reliable ages are not available for HAT-P-7 and WASP-1, but their low \vsini\footnote{The \vsini\ values reported in Table~\ref{tab:comp} should be considered upper limits as \citet{torres2012} ignored possible contributions from ``macroturbulence''.} and \logR\ values indicate they certainly are not young.
Distances were estimated based on the $V$-band magnitudes, \Teff, and radii; reddenings E($B-V$) then were derived from the extinction maps of \citet{amores2005}.
\clearpage
\clearpage
\begin{table}
\caption{\label{tab:comp} Fundamental parameters of WASP-12 and two comparison stars.}
\begin{tabular}{lcccccc}
\tableline\tableline
Star & \Teff & \vsini & [Fe/H] & \logR & distance & E($B-V$) \\
name & [K]   & [\kms] & [dex]  &       & [pc]     & [mag]    \\
\tableline
HAT-P-7	  & 6441 &  5.0 & $+$0.15 & $-$5.018 & 310 & 0.027 \\
WASP-1	  & 6160 &  1.7 & $+$0.14 & $-$5.114 & 388 & 0.020 \\
\tableline
WASP-12	  & 6250 & $<$1 & $+$0.28 & $-$5.500 & 439 & 0.144 \\
\tableline
\end{tabular}
\tablecomments{\Teff, \vsini, and metallicity values are from \citet{torres2012}; \logR\ values are from \citet{knutson2010}; distances and reddening were derived as explained in the text.}
\end{table}

Because WASP-12 has the lowest \vsini\ value, we broadened its spectrum to match the comparison tracings. Flux normalisation was performed in two steps. We removed the blaze function adjusting the continuum using a synthetic spectrum with atmospheric parameters appropriate to the three stars. We then matched the fluxes of the far line wings to WASP-12, for the H and K lines separately, using the windows indicated in Fig.~\ref{fig:corot1wasp1}. The latter adjustments were all $\leq$5\%.

As anticipated from the earlier \ion{Mg}{2}\,h\&k line observations \citep{haswell2012}, the \ion{Ca}{2}\,H\&K line cores of WASP-12 are significantly deeper than those of HAT-P-7 and WASP-1, despite similarities in the stellar properties. The anomaly covers a $\sim$4\,\AA\ wide region, similar to the \ion{Mg}{2}\,h\&k depression. The same comparison relative to CoRoT-1, HAT-P-8, HAT-P-16, HAT-P-24, HAT-P-33, HAT-P-39, Kepler-8, Kepler-25, TrES-4, and WASP-3, always shows the central $\sim$4\,\AA\ of the line core the most depressed in WASP-12. The flux adjustment might seem to be challenging in this heavily blanketed spectral region, but there are enough relatively clean intervals in the \ion{Ca}{2} wings that our results are very little affected by the normalization procedure. We carefully checked that background light under-/over-correction could not explain the observed anomaly in the WASP-12 line core.

\begin{figure}
 \includegraphics[width=17cm]{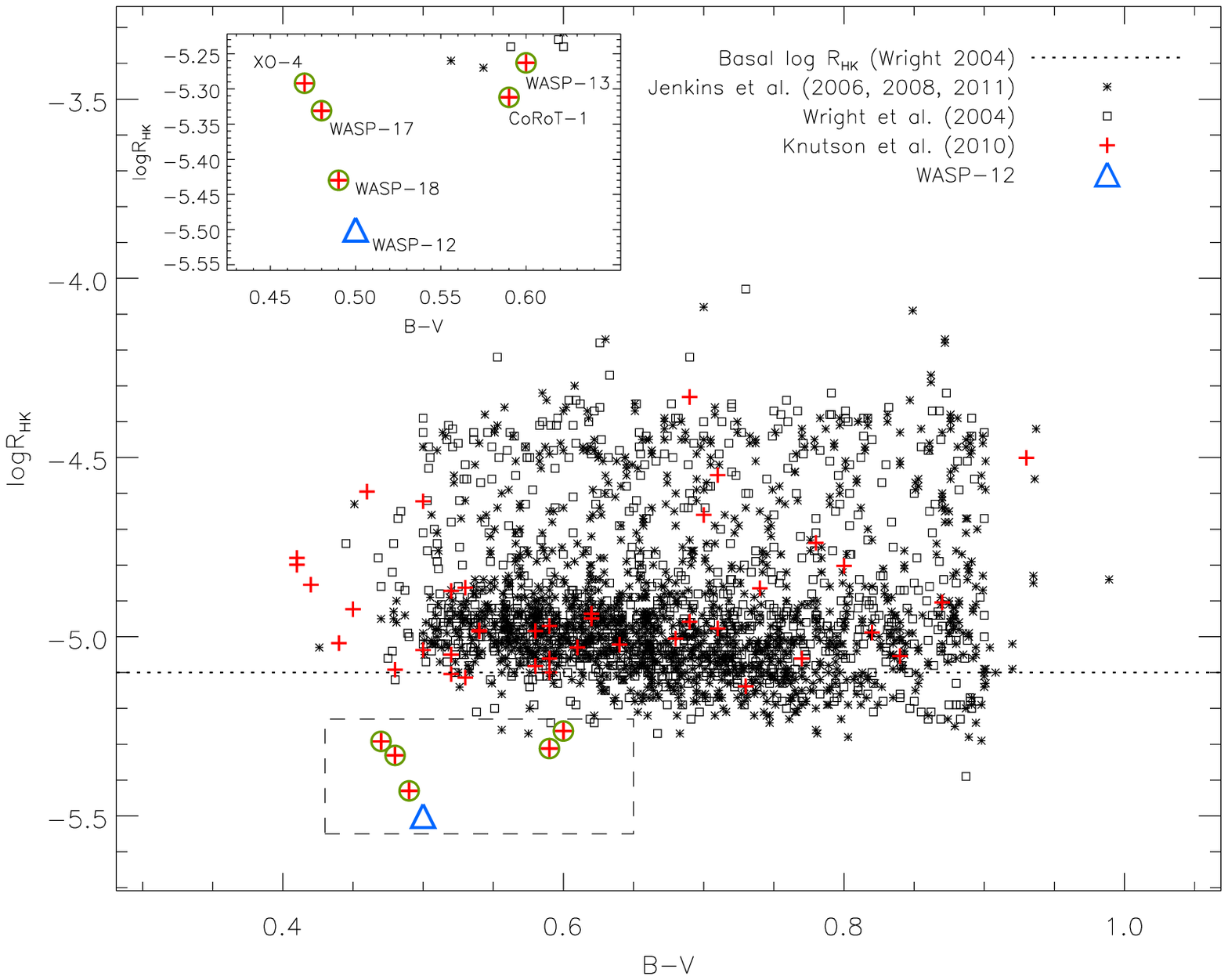}
\caption{\label{fig:logR} WASP-12 (blue triangle) in $B-V$ vs. \logR\ plane, compared with stars with $B-V\,<\,1.0$ observed by \citet{wright2004} (squares), \citet{jenkins2006,jenkins2008,jenkins2011} (stars), and \citet{knutson2010} (red pluses). Color and activity indices adopted for WASP-12 are those given by \citet{knutson2010}. Circles indicate  positions of planet hosting stars X0-4, CoRoT-1, WASP-13, WASP-17, and WASP-18 (see Sect.~\ref{discussion}). The dotted line indicates the minimum activity value within the \citet{wright2004} sample, accounting for contamination by subgiants \citep{wright2004a}.}
\end{figure}

\citet{knutson2010} measured the \logR\ activity index in 50 transiting planet host stars, finding that WASP-12 was the least active, by that measure, with a remarkably low \logR\,=\,$-5.500$. Figure~\ref{fig:logR} shows that  WASP-12 is an extreme outlier
in a color ($B-V$) vs. activity (\logR) plane populated by the samples from \citet{wright2004}, \citet{jenkins2006,jenkins2008,jenkins2011}, and \citet{knutson2010}. As the \logR\ index is logarithmic, WASP-12 indeed  falls well below the lower boundary of the envelope of activity displayed by the stars considered here. For solar metallicity dwarfs the lower bound is at $-5.1$ \citep{henry1996,wright2004a} and it decreases slightly to $-5.15$ for stars with the somewhat enhanced metallicity of WASP-12 \citep{saar2011}.
The lower bound of the index results from a ``basal'' level of chromospheric emission which seems to depend mainly on \Teff\ and only weakly on metallicity (as noted above). Stars with indices above the basal level are affected by varying degrees of dynamo magnetic activity, which does correlate with the rotational velocity. The basal flux level should be a hard lower bound on the chromospheric emission of a late-type main sequence star, and any otherwise normal dwarf that falls below the basal level must be affected by extrinsic absorption which attenuates the intrinsic core emission.
To be sure, evolved subgiants can fall below the basal boundary, as discussed by \citet{wright2004a}. Figure~\ref{fig:logR} contains contamination by such subgiants, but they can be discriminated by the broadening of their \ion{Ca}{2} emission cores by the ``Wilson-Bappu effect'' \citep[e.g.,][]{ayres1979}. In short, as suggested in \citet{haswell2012}, it is likely that the abnormally low \logR\ index of WASP-12 derived by \citet{knutson2010} is simply biased by extrinsic absorption, hiding the chromospheric signatures that must be present intrinsically in \ion{Ca}{2}\,H\&K.
\clearpage
\section{ISM measurements along the WASP-12 sightline}\label{ism}
To determine the contribution of interstellar absorption to the depressions in the cores of both \ion{Mg}{2} and \ion{Ca}{2}, we carried out a study of the ISM column in the WASP-12 sightline. We used  \espa\ \citep{donatietal1997} at the Canada-France-Hawaii Telescope (CFHT) and  FIES  at the Nordic Optical Telescope (NOT) to obtain high resolution spectra of three early-type stars (HD257926, HD258049, HD258439) lying within 20$\arcmin$ of WASP-12, and at about the same distance. We observed hot stars for three reasons: 1) they are  brighter than WASP-12 at the same distance, allowing high-S/N measurements; 2) they are typically  fast rotators, simplifying identification of the sharp ISM absorptions; and 3) they do not exhibit stellar activity, which might otherwise affect interpretations of the \ion{Ca}{2}\,H\&K lines.

The \espa\ spectra cover 3700--10400\,\AA, with a resolving power of 80\,000 in the ``star only'' instrumental configuration, while the FIES spectra cover 3700--7400\,\AA, with a resolving power of 67\,000. The spectra revealed that HD258049 is a spectroscopic binary, while HD258439 is a chemically peculiar star (possibly magnetic), and therefore less straightforward to model. For this reason we focused on HD257926, apparently a chemically normal early A-type star. It lies 15.7$\arcmin$ away from WASP-12 on the sky, at about twice the distance ($\sim$650\,pc). We estimated the fundamental parameters by comparing observed hydrogen line profiles (H$\alpha$, H$\beta$, H$\gamma$, and H$\delta$) with synthetic spectra calculated with the \synth\ code \citep{kochukhov2007}, utilizing \llm\ model atmospheres \citep{shulyak2004}. We obtained: \Teff$\sim$10700\,K, \logg$\sim$4.25, \vsini\ $\sim$120\,\kms.

\begin{figure}
\includegraphics[width=8.5cm]{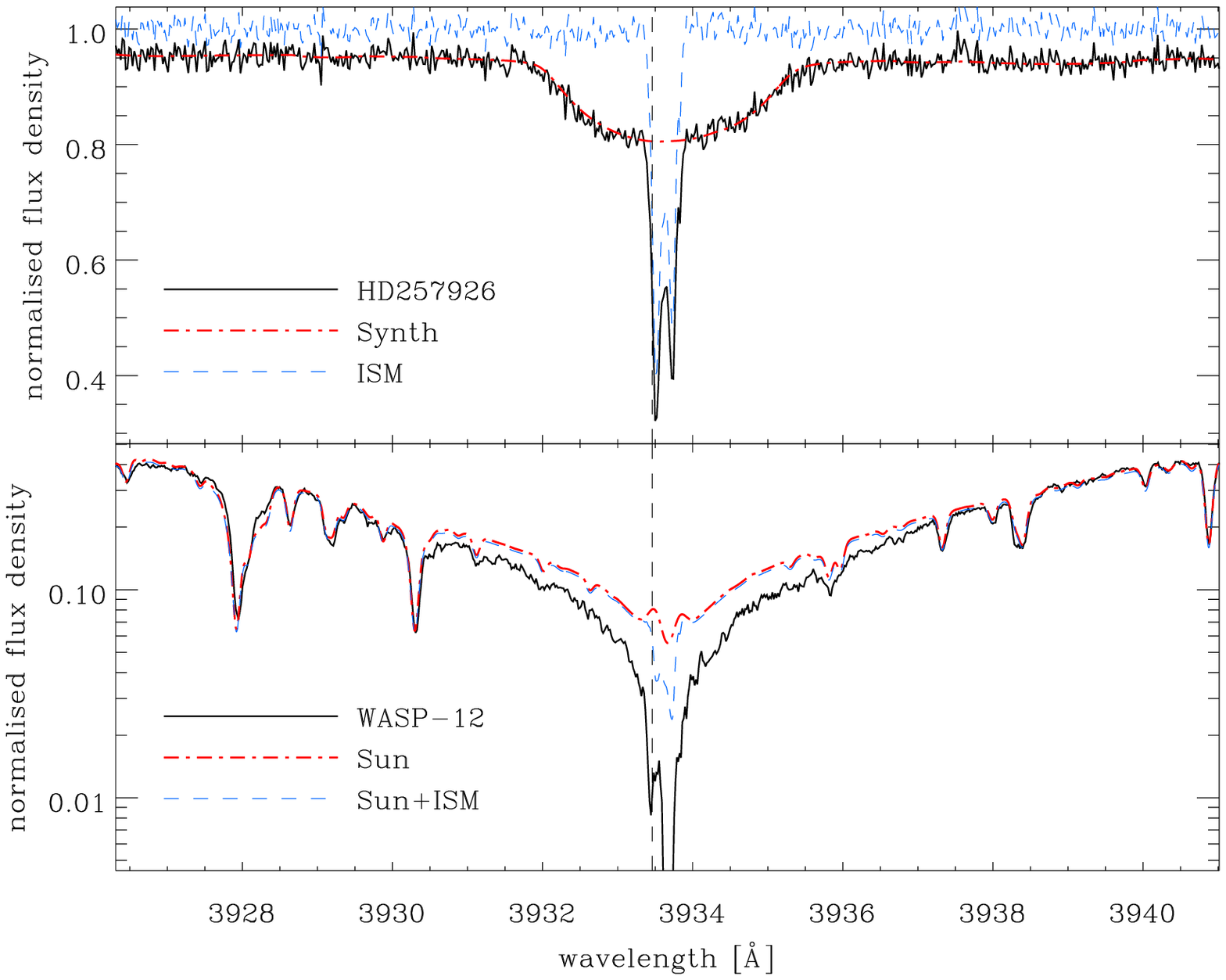}
\includegraphics[width=8.5cm]{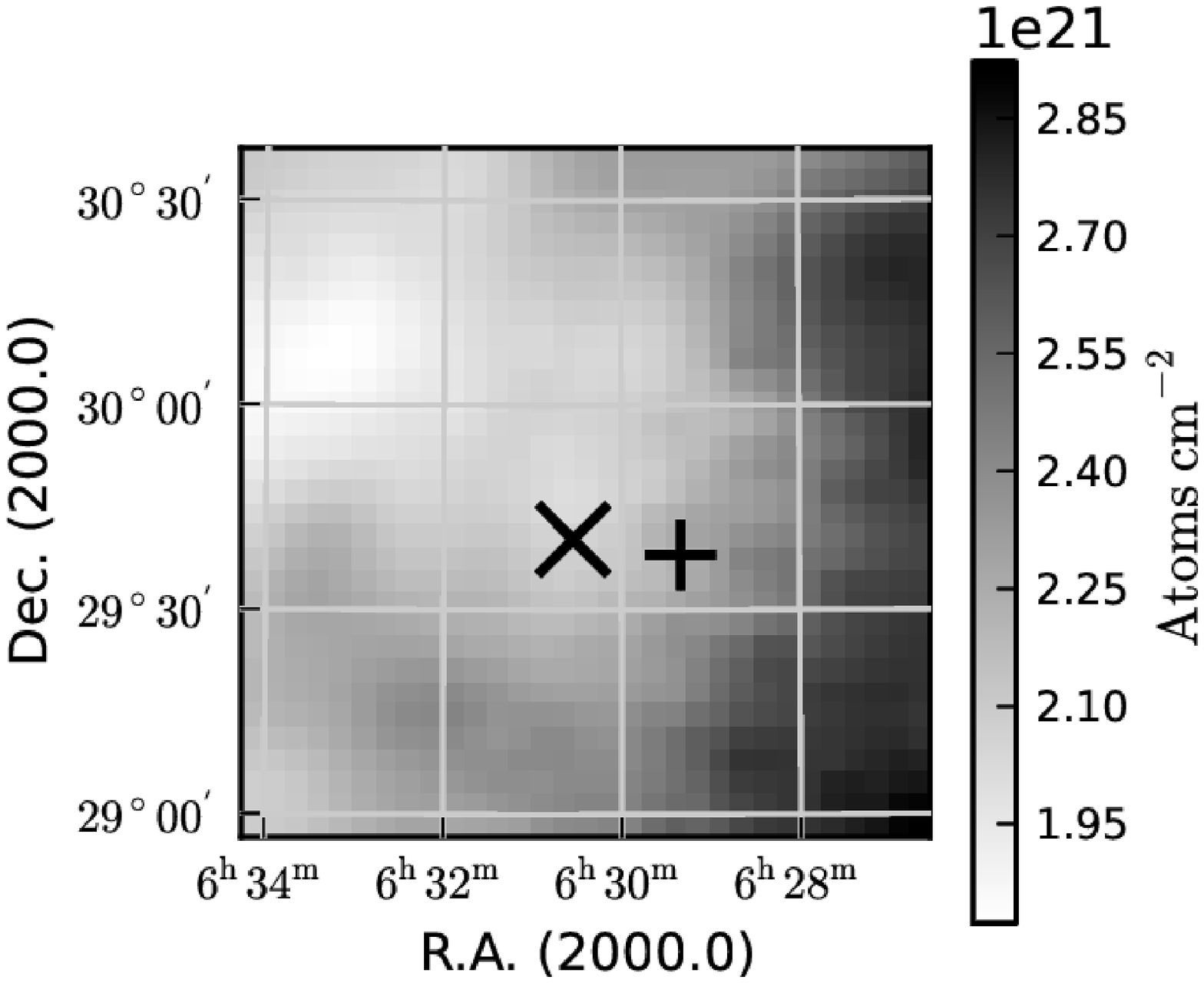}
\caption{\label{fig:CaNa} Top-left panel: HD257926 (solid line), observed with \espa; synthetic spectrum (dash-dotted line) used to recover the ISM \ion{Ca}{2}\,K absorption profile (dashed line). Bottom-left panel: WASP-12 (solid line) compared to the Sun (dash-dotted line: National Solar Observatory Digital Archive). The Solar spectrum was normalised with the procedure of Fig.~\ref{fig:corot1wasp1}, and convolved to match the HIRES resolution. Note the excellent agreement in the far wings. The dashed line applies ISM absorption obtained from HD257926 to the Solar spectrum. The dashed vertical line indicates the position of the ISM-unrelated feature in WASP-12's spectrum. Right panel: EBHIS total \ion{H}{1} column density map in the WASP-12's direction. Cross and plus mark the positions of WASP-12 and HD257926, respectively.}
\end{figure}

Narrow ISM absorption lines are clearly  visible in HD257926's  \ion{Ca}{2}\,H\&K and \ion{Na}{1}\,D1 and D2 lines: see Fig.~\ref{fig:CaNa} (top-left panel) for \ion{Ca}{2}\,K. Each doublet member shows at least two ISM velocity components. For conciseness we focus on the higher S/N \espa\ spectrum, but the FIES spectrum provided comparable results. We removed the stellar contribution using a synthetic spectrum, and fitted multiple Gaussians to the residual ISM absorptions. Table~\ref{tab:ism} lists the equivalent widths of the ISM lines. We calculated synthetic ISM line profiles for a range of ``$b$'' parameters (characterizing thermal and turbulent line broadening) and column densities, and for each $b$ value determined the column density that reproduced the observed equivalent width. For \ion{Ca}{2}, we then compared the derived H and K column densities, and determined the $b$ value for which the two column densities agreed. We proceeded in the same way for the \ion{Na}{1}\,D lines. In all cases we allowed the two Gaussian profiles to have variable widths. Our conclusions were not significantly affected by these choices.
\clearpage

Uncertainties were based on (S/N)$^2$ values of the \espa\ spectrum of HD257926, separately in the regions of the \ion{Ca}{2} and \ion{Na}{1} lines. We then applied a Monte Carlo scheme to estimate the parameter uncertainties in the Gaussian fitting. 

In estimating the final \ion{Ca}{2} and \ion{Na}{1} column densities, we considered the effect of the equivalent width uncertainties and the influence of the $b$ parameter, which was varied in the 2.8--4.0\,\kms\ range, compatible with the spread of values observed in the direction of WASP-12 \citep{welsh2010}. For the \ion{Ca}{2} lines, we derived $b$ values of $\sim$3.5\,\kms, while for the \ion{Na}{1} lines we derived $b$ values of $\sim$4 and $<$3\,\kms\ for the weak and strong components, respectively. The smaller $b$ value of the stronger \ion{Na}{1} components suggests a lower temperature in the ISM clouds producing these absorptions.

Table~\ref{tab:ism} summarises the final results. The column densities and $b$ parameters derived for \ion{Ca}{2} and \ion{Na}{1} are consistent with those reported by \citet{welsh2010} for stars at a distance of $\sim$400\,pc in a direction close to that of both HD257926 and WASP-12. The firm conclusion, as illustrated graphically in the lower panel of Fig.~\ref{fig:CaNa}, is that the ISM absorption along the WASP-12 sightline is insufficient to cause the apparent broad absorption seen in the \ion{Mg}{2} and \ion{Ca}{2} cores. Figure~\ref{fig:CaNa} also shows the position of a narrow absorption feature in WASP-12's line profile, which is absent from the ISM absorption. It is blue-shifted by 14.5 and 14.2\,\kms\ for the \ion{Ca}{2}\,H\&K lines, respectively. This feature is absent in all the other stars we examined in Sect.~\ref{ca}.

Nevertheless ISM absorption might vary over small angular distance, but the total \ion{H}{1} column density maps from the Effelsberg-Bonn HI Survey \citep[EBHIS;][]{hi1,hi2} show (Fig.~\ref{fig:CaNa}) that the \ion{H}{1} column density variation within 20$\arcmin$ from WASP-12 is small and WASP-12 has a lower column than HD257926.

\clearpage
\begin{table}
\caption{ISM properties derived from HD257926}
\begin{tabular}{lcccc}
\tableline\tableline
 & \ion{Ca}{2}\,K & \ion{Ca}{2}\,H & \ion{Na}{1}\,D1 & \ion{Na}{1}\,D2 \\
\tableline
\vr$_{-blue}$  &  $+$1.8$\pm$0.1 &  $+$2.0$\pm$0.2 &  $+$4.2$\pm$0.2 &  $+$4.2$\pm$0.2 \\
\vr$_{-red}$   & $+$17.3$\pm$0.2 & $+$18.0$\pm$0.2 & $+$17.0$\pm$0.1 & $+$16.9$\pm$0.1 \\
FWHM$_{blue}$  &    10.0$\pm$0.2 &     9.8$\pm$0.5 &     8.0$\pm$0.5 &    10.5$\pm$0.5 \\
FWHM$_{red}$   &    11.7$\pm$0.3 &    10.4$\pm$0.7 &     9.3$\pm$0.2 &    10.1$\pm$0.2 \\
EQW$_{blue}$   & 82$\pm$2        & 56$\pm$2        &  35$\pm$2       &  79$\pm$4       \\
EQW$_{red}$    & 79$\pm$2        & 48$\pm$3        & 182$\pm$2       & 212$\pm$3       \\
\tableline
log\,N$_{blue}$   & \multicolumn{2}{c}{12.2$\pm$0.1} & \multicolumn{2}{c}{11.6$\pm$0.1} \\
log\,N$_{red}$    & \multicolumn{2}{c}{12.1$\pm$0.1} & \multicolumn{2}{c}{12.8$\pm$0.1} \\
\tableline
log\,N$_{SS96}$   & \multicolumn{2}{c}{9.79}  & \multicolumn{2}{c}{11.37} \\
log\,N$_{Sun}$  & \multicolumn{2}{c}{13.52} & \multicolumn{2}{c}{12.25} \\
\tableline
\end{tabular}
\tablecomments{\label{tab:ism} Radial velocity (\vr\ - in \kms), full width at half maximum (FWHM - in \kms), equivalent width (EQW - in m\AA), and column density (in cm$^{-2}$) of the \ion{Ca}{2}\,H\&K and \ion{Na}{1}\,D1 and D2 ISM velocity components (``blue'' and ``red''), measured in HD257926. The final two rows list total \ion{Ca}{2} and \ion{Na}{1} column densities derived for average ISM conditions and the WASP-12 reddening, assuming either the gas-phase ISM abundances of Savage \& Sembach~(1996 - SS96) or the solar abundances of \citet{asplund2009} (see text).}
\end{table}

Table~\ref{tab:ism} also lists column densities obtained assuming average interstellar conditions, N$_H$=5.8$\times$10$^{21}$$\times$E($B$-$V$) \citep{savage1979}, the ISM ionization mixing by \citet{frisch2003}, and either the gas-phase ISM abundances of \citet{savage1996} or the solar abundances of \citet{asplund2009}. The measured Na and Ca ISM column densities in the direction of WASP-12 apparently are slightly larger than the average ISM, compatible with the presence of a relatively dense \ion{H}{1} region in that direction \citep{ben2012}. We estimated the corresponding \ion{Mg}{2} column density assuming the solar abundance, obtaining log\,N=16.42 in agreement with the value proposed by \citet{haswell2012}, but far below the log\,N=17.30 required to reproduce the full \ion{Mg}{2} line core depressions \citep{haswell2012}.
\section{Discussion}\label{discussion}
In Sections~\ref{ca} and \ref{ism} we showed that neither lack of activity nor strong ISM absorption is likely to be responsible for the 
anomalously deep resonance line cores in  WASP-12. \citet{haswell2012} suggested that the anomaly might be caused by extrinsic absorption by diffuse material within the WASP-12 planetary system.
\citet{fossati2010b} searched for the infrared signature of a cool component to WASP-12's SED, but found none.

It is well established that WASP-12\,b is experiencing a blow-off phase \citep[e.g.][]{li2010,fossati2010a,haswell2012}. Thus, the planetary atmosphere would be an obvious reservoir of gas to fill a circumstellar disk/cloud.
Given the variability of chromospheric activity and that the planet's orbit is only one stellar diameter away from the surface of WASP-12, there are many processes stripping gas from the planet, and subsequently acting upon it, making the geometry complex and time variable. The broadband NUV transits reported by \citet{haswell2012} directly revealed dramatic variability in the diffuse gas distribution.

\citet{li2010} suggested that gas lost by the planet would fall toward the star, forming an accretion disk interior to the planet. In contrast, \citet{vidotto2010} interpret the observations in terms of bow-shocked material entrained in the planet's magnetosphere. It is certain that the stellar wind and the stellar radiation pressure, along with the dynamic magnetic fields of both planet and star, play important roles in determining the
time-dependent behaviour of the gas. The diffuse gas, which shares the planet's specific angular momentum, but has a greatly enhanced specific cross-section to radiation, is likely to move outwards from the planet's orbit as suggested in \citet{haswell2012}. In that scenario, it is easy to imagine that the gas will not be tightly constrained to the orbital plane, but will disperse to cover much of the disk of the star from our near-edge-on viewing angle. In the accretion disk scenario of \citet{li2010} it is less obvious that the diffuse gas will shadow the widespread bright chromospheric ``network'' on the stellar disk, but the simulation by \citet{bisikalo2013}, suggests that the planet could expel gas to distances as large as five planet radii perpendicular to the planet's orbit (about half the stellar diameter, if the ejection is symmetrical about the orbital plane). The relatively high temperature of the disk should prevent it settling onto the orbital plane. But, if this indeed were to occur, the large mass of the material comprising the disk would force the settling time scale to be longer than the orbital period, therefore the planet could continually pump up the disk thickness by feeding new material into it.

We have argued that the flux depression anomaly we detected in the cores of the \ion{Mg}{2} and \ion{Ca}{2} resonance lines of WASP-12 is evidence for the presence of a translucent circumstellar disk/cloud of material lost by the evaporating exoplanet. Figure~\ref{fig:logR} (bottom-left corner and insert) highlights five other stars, also hosting hot-Jupiters, that have anomalously low stellar activity indices: X0-4, CoRoT-1, WASP-13, WASP-17, and WASP-18. The latter is particularly noteworthy given the many similarities with the WASP-12 system. The only significant differences are the planet mass (WASP-18\,b is $\sim$7 times more massive than WASP-12\,b) and the stellar age \citep[WASP-18 is only $\sim$0.6\,Gyr old;][]{brown2011}. The heavier companion and larger stellar activity would undoubtedly have an impact on the geometry and stability of a planet-fed disk/cloud. A practical advantage is that WASP-18 is 2.3 magnitudes brighter than WASP-12, allowing more precise measurements of the planetary exospheric radius, as well as the prospect of acquiring far-UV spectra (which would conclusively settle the ``activity'' issue). Our findings and these considerations make the WASP-18 planetary system a key target for future UV observations to continue to build the legacy of HST studies of evaporating hot-Jupiters.
\acknowledgments
This work is based on observations obtained at the Canada-France-Hawaii Telescope and Nordic Optical Telescope.
We thank the CFHT staff, Dr. Nadine Manset, Dr. Daniel Devost, Dr. Doug Simons, for their help in obtaining the observations. Astronomy research at the Open University is supported by an STFC rolling grant (CH). LF acknowledges support by the International Space Science Institute in Bern, Switzerland and the ISSI team ``Characterizing Stellar and Exoplanetary Environments''. The EBHIS (Effelsberg-Bonn HI Survey) is funded by the Deutsche Forschungsgemeinschaft under grants KE757/7-1 to 7-3. EBHIS is based on observations with the 100-m telescope of the MPIfR (Max-Planck-Institut f\"ur Radioastronomie) at Effelsberg.



{\it Facilities:} \facility{CFHT (ESPaDOnS)}, \facility{NOT (FIES)}, \facility{KECK (HIRES)}, \facility{Radio Telescope Effelsberg}

\end{document}